\documentstyle[12pt]{article}

\parskip=\medskipamount            
\def\7#1#2{\mathop{\null#2}\limits^{#1}}        

\def\beee{\begin{equation}}
\def\eeee{\end{equation}}
\def\dggg{^{\dagger}}

\def\dels {{\nabla \hspace{-8.7pt} \slash}\;}
\def\QQ {{p \hspace{-4.5pt} \slash}\;}

\begin{document}

\bibliographystyle{unsrt}
\begin{center}
\textbf{PATH INTEGRALS FOR PARASTATISTICS}\\
[5mm]
O.W. Greenberg\footnote{email address, owgreen@physics.umd.edu.}
and A.K. Mishra\footnote{Permanent address
Institute of Mathematical Sciences, Chennai 600 113, India,
email address: mishra@imsc.res.in }\\

{\it Center for Theoretical Physics\\
Department of Physics \\
University of Maryland\\
College Park, MD~~20742-4111}\\
~\\
University of Maryland Preprint PP 04-044\\
~\\
\end{center}

\begin{abstract}

We demonstrate that parastatistics can be quantized using path integrals
by calculating the generating functionals for time-ordered products
of both free and interacting parabose and parafermi fields in terms of path
integrals. We also give a convenient form of the commutation relations for
the Green components of the parabose and parafermi operators in both the
canonical and path integral formalisms.

\end{abstract}

\section{Introduction}
Canonical quantization~\cite{hei, wei} and path integral
quantization~\cite{fey,cha} are two widely
used methods to quantize a physical system. The generality of these
two methods and their equivalence or non-equivalence has general interest.
Many physicists, including the present authors, believe that any system
that can be quantized using one of these methods also can be quantized
using the other. We acknowledge that for technical reasons one or the
other method may be simpler or more convenient for any specific system.

Some years ago M.G.G. Laidlaw and C. M. DeWitt~\cite{lai} in
a well-cited influential paper studied path integral quantization for
a system of identical particles. They noted the important distinction
between quantization of identical particles in two space dimensions and
in three or more dimensions. This distinction holds for both path integral
and canonical quantization. We believe that their paper
correctly proves that parastatistics cannot be based on the topology
of three dimensional space (in contrast to anyon statistics~\cite{lei, wil}, 
which is
based on the topology of two dimensional space). 
Although the text of their paper does not
mention parastatistics, their abstract contains the phrase ``... showing
that the Feynman formalism excludes parastatistics.'' Dr. Cecille DeWitt has
informed us that the abstract is misleading and that M. Laidlaw and C. DeWitt
did not intend to state that there is a contradiction between path
integration and parastatistics~\cite{dew}. If, indeed,
parastatistics fields~\cite{gre} could not be quantized using path integrals
that would have significance for the general
question of the equivalence of the canonical and path integral methods
of quantization. Our belief that any field theory that can be quantized
using the canonical formalism also can be quantized using path integrals
was our initial motivation to formulate field theories with parastatistics
fields using path integrals.

There are several plausible assumptions about the quantum mechanics of
indistiguishable particles that lead directly to the conclusion that
only bosons or fermions can exist in three or more space dimensions.
These plausible assumptions lead to the same conclusion in the context of
path integrals. The basic principles of quantum mechanics do not require these plausible
assumptions; indeed without these
assumptions quantum mechanics allows the description of
indistinguishable particles
that have more general statistics than bose or fermi
statistics~\cite{mesowg,hil}.
Here are some of the plausible assumptions together with the reasons why these
plausible assumptions are not required for quantum mechanics: (1) Transposing
indistinguishable particles in a quantum state produces a state that is
linearly dependent on the initial state, i.e. $T \Psi = \eta \Psi$,
where $T$ is a transposition. Since
the square of a transposition is the identity this leads directly to
$T^2 \Psi = \eta^2 \Psi =\Psi$, so $\eta = \pm 1$. In a general representation
of the symmetric group this is not the case. Greenberg and Messiah showed~\cite{mesowg}
that quantum mechanics can be formulated for an arbitrary represention of the 
symmetric group. (2) The space of
quantum states obtained by permuting
indistiguishable particles is coherent, i.e., these states can be superposed
to form other states without restriction. For states of several
indistinguishable particles there is an absolute selection rule that prohibits
transitions between states in inequivalent
irreducible representations of the symmetric group so that the relative
phases of states in inequivalent
irreducible representations can never be measured. Such
systems must be treated as statistical mixtures rather than as
coherent superpositions of states. In short, there is a superselection rule
that separates states in inequivalent
irreducible representations of the symmetric group~\cite{mesowg,ama}.
If one assumes all states
can be superposed, then since there is a symmetric state and an antisymmetric
state of $N$ particles for each $N$, only bosons and fermions can occur.
(3) The configuration space of a system of particles is in one-to-one
correspondence with the states of the system. This assumption implies that
the states always can be superposed and leads back to assumption (2).
(4) The homotopy classes of the configuration space determine the particle
statistics. The discussion of irreducible representations of the symmetric group
shows that the existence of different homotopy classes allows different
types of statistics, but there is no requirement that each type of statistics
must be associated with a different homotopy class of the configuration
space.

\section{Canonical Quantization of Parastatistics}
H.S. Green~\cite{gre} proposed the first proper quantum statistical
generalization of bose and fermi statistics.  Green noticed that the commutator
of the number operator with the annihilation and creation operators is the same
for both bosons and fermions
\beee
[n_k, a\dggg_l]_-=\delta_{kl}a\dggg_l.                 \label{number}
\eeee
The number operator can be written
\beee
n_k=(1/2)[a\dggg_k, a_k]_{\pm}+ {\rm const.},
\eeee
where the anticommutator $[a\dggg_k, a_k]_+$
(commutator $[a\dggg_k, a_k]_-$) is for the bose (fermi) case.  If these
expressions are inserted in the number operator-creation operator commutation
relation, the resulting relation is {\it trilinear}
in the annihilation and creation operators.  Polarizing the number operator to
get the transition operator $n_{kl}$ which annihilates a free particle in state
$l$ and creates one in state $k$ leads to Green's trilinear commutation relation
for his parabose and parafermi statistics,
\beee
[[a\dggg_k, a_l]_{\pm}, a\dggg_m]_-=2\delta_{lm}a\dggg_k            \label{paracr}
\eeee
Since these rules are trilinear, the usual vacuum condition,
\beee
a_k|0\rangle=0,
\eeee
does not suffice to allow calculation of matrix elements of the $a$'s and
$a\dggg$'s; a condition on single-particle states must be added,
\beee
a_k a\dggg_l|0\rangle=p \delta_{kl}|0\rangle.             \label{norm}
\eeee

Green found an infinite set of solutions of his commutation rules, one for each
positive integer $p$, by giving an ansatz which he expressed in terms of
bose and fermi operators.  Let the annihilation and creation operators be
represented by a sum of operators with an additional index, the ``Green'' index,
\beee
a_k\dggg=\sum_{\alpha=1}^p b_k^{(\alpha) \dagger},~~a_k=\sum_{\alpha=1}^p b_k^{(\alpha)},
\eeee
and let the $b_k^{(\alpha)}$ and $b_k^{(\beta) \dagger}$
be bose (fermi) operators
for $\alpha=\beta$ but anticommute (commute) for $\alpha \neq \beta$ for the
``parabose'' (``parafermi'') cases.  This ansatz clearly satisfies Green's
relation since the (polarized) number operator is diagonal in the Green index for
each case.  The integer $p$ is the order of the parastatistics.  The physical
interpretation of $p$ is that, for parabosons, $p$ is the maximum number of
particles that can occupy an antisymmetric state, while for parafermions, $p$
is the maximum number of particles that can occupy a symmetric state (in
particular, the maximum number which can occupy the same state).  The case $p=1$
corresponds to the usual bose or fermi statistics.
Later, Greenberg and Messiah~\cite{mesowg} proved that Green's ansatz gives
all Fock-like solutions of
Green's commutation rules.

We give a useful form of the commutation relations for the Green components
of a parabose (parafermi) operator,
\beee
b_k^{(\alpha)} b_l^{(\beta) \dagger} = \pm (2 \delta_{\alpha,\beta} -1)
b_l^{(\beta) \dagger}b_k^{(\alpha)} + \delta_{\alpha, \beta} \delta_{k,l},    \label{delta}
\eeee
where the upper (lower) sign is for parabose (parafermi);
analogous forms hold when both operators are annihilation or both are
creation operators, but the $\delta_{\alpha, \beta} \delta_{k,l}$ term is absent.
As far as we know this form does not appear in the literature. We find that
the $\pm (2 \delta_{\alpha,\beta} -1)$ factors are very useful in doing systematic
calculations.

We discussed quantization of parastatistics above in terms of the commutation
relations for parabose and parafermi creation and annihilation operators.
Quantization of parastatistics in terms of spacetime-dependent field
operators follows for free fields or fields in the interaction picture by
the usual Fourier transformations using, for example, plane waves. Thus for
a neutral scalar parabose field,
\beee
\phi(x)=\frac{1}{(2 \pi)^{3/2}}\int
\frac{d^3k}{\sqrt{2 \omega_k}}(a_k e^{-i k \cdot x} +
a\dggg_k e^{i k \cdot x}),~\omega_k=\sqrt{{\mathbf{k}}^2+m^2}.
\eeee
In terms of fields, local observables have a form analogous to the usual
ones; for example, the local current for a spin-1/2 parafermi theory is
\beee
j_{\mu}=(1/2)[\bar{\psi}(x),\gamma_{\mu} \psi(x)]_-=
\sum_{\alpha=1}^p \bar{\psi}^{(\alpha)}(x)\gamma_{\mu} \psi^{(\alpha)}(x).
\eeee
  From Green's ansatz, it is clear
that the squares of all norms of states are positive, since sums of bose or
fermi operators give positive norms.  Thus parastatistics gives a set of
orthodox theories.

We give the form of the field commutation relations for parabose
and parafermi fields in terms of both the the fields and the Green
components of the fields,
\beee
[[\phi(x_1), \phi(x_2)]_{\pm}, \phi(x_3)]_-=2i\Delta(x_2-x_3)\phi(x_1)
\pm 2 \phi(x_2) i\Delta(x_1-x_3)
\eeee
\beee
\phi^{(\alpha)}(x_1) \phi^{(\beta)}(x_2)=(2\delta_{\alpha, \beta}-1)
 \phi^{(\beta)}(x_2) \phi^{(\alpha)}(x_1) +\delta_{\alpha, \beta} i \Delta(x_1-x_2)
\eeee
For charged fields, keep only the contractions between a field and its adjoint.
 
\section{Green Ansatz for Path Integrals}
\subsection{Parabose case}
Since parafields can be constructed
in second quantized form using the Green ansatz~\cite{gre}, we
have constructed the representations in terms of path integrals
by using constructions closely analogous to those of Green. Since,
as mentioned above,
Greenberg and Messiah~\cite{owgmes} proved that any field
theory that obeys the trilinear commutation relation of Green that has
a Fock-like representation can be represented using the Green ansatz,
the Green ansatz for the path integral can be used without loss of
generality.

We recall the generating a (bose) charged scalar field given
in terms of a path integral,
\begin{eqnarray}
\lefteqn{W[J,J^{\star} ] = }     \nonumber    \\
& & \int {\cal D} \phi^{\star}~ {\cal D} \phi ~
 \exp \left[ i \int d^4x \left( {\cal L}(\phi(x),~\phi^
\star (x)) + J^\star (x) \phi(x) + \phi^{\star}(x) J(x) \right)\right]
                                     \label{gen}
\end{eqnarray}
with
\beee
{\cal L} = {\cal L}_0+{\cal L}_I,~{\cal L}_0=
\partial_\mu \phi^\star  \partial^{\mu} \phi - m^2 \phi^\star
 \phi,~{\cal L}_I= V (\phi, \phi^\star).          \label{kg}
\eeee
Since one can calculate the path integral for the free term ${\cal L}_0$
explicitly, one often represents the interaction term ${\cal L}_I$ by a
functional derivative. Then
\begin{eqnarray}
W[J,J^{\star}] &=&
W[0,0] \exp[i \int d^4x
V(-i \frac{\delta}{\delta J^{\star}(x)}, -i \frac{\delta}{\delta J(x)})]
\nonumber \\
 & & \times~ \exp[-i
\int d^4x d^4y J^\star (x) \Delta_F(x-y) J(y)],                \label{int}
\end{eqnarray}
where $\Delta_F(x)=(2\pi)^{-4}\int d^4k
\exp(-ik \cdot x)(k^2 - m^2 -i \epsilon)^{-1}$.

We construct the generating functional for time ordered products for parabose
fields in parallel with the above construction for bose fields. We must take into
account the requirement that the terms allowed in the parabose (and also in the
parafermi) Lagrangian must obey local commutativity~\cite{owgmes, agt}. These
papers show that the para fields must occur in one of two forms.
The form
$[\phi^{\star}(x),\phi(x)]_{\pm}$ for any order $p$ of parastatistics is
analogous to the requirement that charged fields occur in the combination
$\phi^{\star}(x)\phi(x)$ in order to obey a $U(1)$ symmetry. The form
of a nested set of $p-1$
commutators and anticommutators for para fields of order $p$ is analogous to
the possibility of a term
$\epsilon^{\alpha,\beta,\cdots, \zeta}\phi_{\alpha}(x) \phi_{\beta}(x)
\cdots \phi_{\zeta}(x)$ in a theory
that obeys $SU(n)$ symmetry. We refer to the
papers just cited for a demonstration and detailed analysis of these possibilities.
In the first case the admissible terms in the Lagrangian will be functions of
terms diagonal in the
Green index. In the second case the admissible terms will be functions of
terms antisymmetric in the Green index;
that is for a given value of $p$ each index from
$1$ to $p$ will occur exactly once in each term.

The Lagrangian density for a free parabose field is
\beee
{\cal L}(\phi, \phi^\star) = \frac {\displaystyle 1}{\displaystyle 2}
([\partial_\mu \phi^\star (x), \partial^\mu \phi(x)]_+  - m^2 [ \phi^\star (x),
 \phi(x)]_+).                        \label{lag}
\eeee
Using the Green ansatz, the parabose field is  represented  by
\beee
\phi (x)
= \sum_{\alpha = 1}^p \phi^{(\alpha)} (x),              \label{pb}
\eeee
that is, as a sum of $p$ bose
fields ($p$ is the order of the parastatistics). The bose fields with
different Green index anticommute rather than commute.
In order to obtain the generating functional,
we introduce parabose Grassmann fields $J(x)$~\cite{eff}. The parabose
Grassmann fields have a representation similar to that for the parabose
fields, except that the right-hand sides always vanish,
\beee
J(x) =\sum_{\alpha = 1}^p J^{(\alpha)}(x),
\eeee
\beee
[J^{(\alpha)}(x),J^{(\beta)}(y)]_{q_{\alpha \beta}}=0,
\eeee
where $q_{\alpha \beta } = 1  -2\delta_{\alpha \beta } $.
The parabose Grassmann Green components commute with the parabose field
Green components for the same value of the Green index and anticommute
for different values of the Green index.
The external source term
$\int d^4 x (1/2)([J^{\star}(x),\phi(x)]_+ + [J(x), \phi^{\star}(x)]_+)$
is then an effective bose operator~\cite{eff}. Using the fact that
\beee
\frac{1}{2}[\partial_\mu\phi^{\star}(x),\partial^\mu\phi(x)]_+ =
\sum_i \partial_{\mu}\phi^{(\alpha)\star}(x)
\partial^\mu\phi^{(\alpha)}(x),    \label{anticom}
\eeee
we write the generating functional for a free parabose field as
\begin{eqnarray}
& &\int \prod_{\alpha=1}^p {\cal D} \phi^{(\alpha)\star} {\cal D}
\phi^{(\alpha)} ~\exp[-i \sum_1^p \int d^4x (
\partial_{\mu}\phi^{(\alpha)\star}(x) \partial^{\mu}\phi^{(\alpha)}(x)
- m^2 \phi^{(\alpha)\star}(x) \phi^{(\alpha)}(x)     \nonumber   \\
& + &  (J^{(\alpha)\star} (x) \phi^{(\alpha)}(x) + h.c.))]. \label{genfun}
\end{eqnarray}
As in the bose case we can do the free path integral to find
\beee
W_{free}[J,J^{\star}] = W_{free}[0,0]
\exp[-i \sum_{\alpha=1}^p
\int d^4x d^4y J^{(\alpha)\star} (x) \Delta_F(x-y) J^{(\alpha)}(y)].   \label{explicitgenfun}
\eeee

For interacting theories in which the terms are functions of
$[\phi^{\star}(x),\phi(x)]_{\pm}$, the interaction term in the Lagrangian
is $V_1([\phi^{\star}(x),\phi(x)]_{\pm})$ and we represent the generating
functional for the interacting theory as
\beee
W[J,J^{\star}]=\exp[i \int d^4x V_1(-\sum_{\alpha=1}^p
\frac{\delta^2}{\delta J^{(\alpha)} \delta J^{(\alpha)\star}})]
W_{free}[J,J^{\star}].
\eeee
We give just one example of the case where the interaction term in the
Lagrangian has the nested commutator form,
\beee
{\cal L_I}=g\frac{1}{4}[[\phi_1(x),\phi_2(x)]_-,\phi_3(x)]_+
=g\sum_{\stackrel{\alpha,\beta,\gamma}{\mathit{all ~different}}}
\phi_1^{(\alpha)}(x)\phi_2^{(\beta)}(x)\phi_3^{(\gamma)}(x)
\eeee
where $\phi_i$ are three different neutral parabose fields of order $p=3$.
In this case we need free Lagrangian terms for each of the three fields,
so the free generating functional is
\beee
W_{free}[J_i,J_j^{\star}] = W_{free}[0,0]
\exp[-i \sum_{i=j=1}^3\sum_{\alpha=1}^3
\int d^4x d^4y J_i^{(\alpha\star)} (x) \Delta_F(x-y) J_j^{(\alpha)}(y)].  \label{freenested}
\eeee
The interacting generating functional is
\begin{eqnarray}
W[J_i,J_j]=\exp[i \int d^3x
\sum_{\stackrel{\alpha,\beta,\gamma}
{\mathit{all ~different}}}
\frac{\delta^3}{\delta J_1^{(\alpha)} \delta J_2^{(\beta)}
\delta J_3^{(\gamma)} }]
W_{free}[J_i,J_j^{\star}].
\end{eqnarray}

Note that for this type of interaction the degree of the fields
in the interaction Hamiltonian must match the order of the parabose
field so that the Green indices are ``saturated.'' In addition,
the interacting Hamiltonian should commute with the field.  For even
degrees the interaction Hamiltonian will anticommute as pointed out
in~\cite{owgmes}.
This case should not be allowed because it
leads to alternating signs of the contribution of the interaction
energy to widely separated states~\cite{eff}.

Now the vacuum matrix elements of the time ordered products of the fields
$\phi^\star (x)$ and $\phi (x)$ are given by the operations
$\sum_1^p \delta / \delta J^{(\alpha)}(x)$ and
$\sum_1^p \delta / \delta J^{(\alpha)\star}(x)$, respectively,
acting on the generating functional.
The anticommutativity of the
$\phi^{(\alpha)}$'s and $J^{(\alpha)}$'s for different values of the Green indices
leads to the anticommutativity of the quantized fields for different
values of the Green index in the vacuum
matrix elements of the time ordered products, and thus leads to the
time ordered products of the parabose fields.

The functional differentiation with respect to $J^{(\alpha)} (x)$ 
can be carried out using the following commutation relation, which is
analogous to Eq.(\ref{delta}) for the canonical formalism,
\beee
\frac {\delta}{\delta J^{(\alpha)} (x)} J^{(\beta)}(y)
= \pm(2 \delta_{\alpha,\beta}-1) J^{(\beta)} (y)
\frac {\delta}{\delta J^{(\alpha)} (x)}+ \delta_{\alpha,\beta}\delta(x-y).     \label{delta2}
\eeee
This relation is for a neutral field and the $\pm$ refers to parabose (parafermi). 
For a charged field, just omit the contractions that do not connect the field with
its adjoint.

As specific illustrations, we evaluate the two-point
and four-point Green's function for a non-interacting parabose system.
We choose to exhibit the results for a neutral field, which has all
terms that can occur in the case of a charged field. For a specific
distribution of a charged field and its adjoint use the result for
the neutral field, but keep only those propagators that connect the
charged field and its adjoint. Discard those propagators that would
have connected the field with the field or the adjoint with the adjoint.
\begin{eqnarray}
<0| T \phi(x_2) \phi (x_1) |0>
&  = & \sum_{\alpha,\beta =1}^p \left(
\frac {\displaystyle 1}{\displaystyle i  } \right)^2
\frac {\displaystyle   \delta^2}{\displaystyle
\delta J^{(\beta)} (x_2) \delta J^{(\alpha)} (x_1)}
W|_{J^{(\beta)} = 0}     \nonumber \\
&  = &   i  p \Delta_F (x_1 - x_2) ,
\end{eqnarray}        \label{2point}
and
\begin{eqnarray}
\lefteqn{<0| T \phi (x_4) \phi (x_3) \phi (x_2) \phi (x_1) |0>
   =     }             \nonumber                         \\
& & \sum_{\alpha,\beta,\gamma,\delta}   \left(
\frac {\displaystyle 1}{\displaystyle i  } \right)^4
\frac {\displaystyle   \delta^4}{\displaystyle \delta J^{(\delta)} (x_4)
 \delta J^{(\gamma)} (x_3) \delta J^{(\beta)} (x_2) \delta J^{(\alpha} (x_1)}
 \times W|_{J = 0}    \nonumber    \\
& = &    p^2 i \Delta_F (x_1 - x_2) i\Delta_F (x_3 - x_4)
+p(2-p) i\Delta_F (x_2 - x_4) i\Delta_F (x_1 - x_3)   \nonumber    \\
& & + p^2 i\Delta_F (x_1 - x_4) i\Delta_F (x_2 - x_3) .
         \label{4point}
\end{eqnarray}

In particular, for a charged field, the result is
\begin{eqnarray}
\lefteqn{<0| T \phi^{\star} (x_4) \phi ^{\star}(x_3) \phi (x_2) \phi (x_1) |0>
   =     }             \nonumber                         \\
& & \sum_{\alpha,\beta,\gamma,\delta}   \left(\frac {1}{ i  } \right)^4
\frac {\delta^4}{\delta J^{(\delta)} (x_4)
 \delta J^{(\gamma)} (x_3) \delta J^{(\beta)\star} (x_2) \delta J^{(\alpha)\star} (x_1)}  
 \times W|_{J, J^{\star}  = 0}    \nonumber    \\
& = &    p(2-p)~ i \Delta_F (x_1 - x_3) i\Delta_F (x_2 - x_4)
 + p^2 i\Delta_F (x_1 - x_4)i \Delta_F (x_2 - x_3).
         \label{4pointcharged} 
\end{eqnarray}

The validity of the above path integral based results can be
ascertained through the direct evaluation of the time ordered products. In momentum
space, the parabose fields are
\beee
\phi (x) = \frac {\displaystyle 1}{\displaystyle (2 \pi)^{3/2}} \int
\frac {\displaystyle d^3 k}{\displaystyle \sqrt{2 \omega_k}}
(b_k e^{-i k.x} + d_k^\dagger e^{i k.x}) \equiv
\phi^{(+)} (x) + \phi^{(-)} (x),    \label{phi}
\eeee
\beee
\phi^\star (x) = \frac {\displaystyle 1}{\displaystyle (2 \pi)^{3/2}} \int
\frac {\displaystyle d^3 k}{\displaystyle \sqrt{2 \omega_k}}
(d_k e^{-i k.x} + b_k^\dagger e^{i k.x}) \equiv
\phi^{\star (+)} (x) + \phi^{\star(-)} (x),     \label{phistar}
\eeee
and these satisfy the vacuum conditions
\beee
\phi^{(+)} (x) |0> =  \phi^{\star (+)} |0> = 0.     \label{vac}
\eeee
Substitution of Eqs.(\ref{phi}, \ref{phistar}) in the time ordered
products, subsequent expansion
of the creation and annihilation operators in terms of Green
components,
and the application of vacuum condition Eq.(\ref{vac}) again lead to
the expressions
Eq.(26) and Eq.(\ref{4point}) for the two and four point
Green's functions. 

\subsection{Parafermi case}
Next we consider the path integral for parafermions.
For a parafermi field we interchange the roles of commutators and
anticommutators relative to the case of parabose fields. Thus the
Lagrangian density for a free parafermi field is
\begin{eqnarray}
{\cal L} (\psi, {\bar{\psi})} & = &  \frac {\displaystyle 1}{\displaystyle 2}
[ \bar{\psi} (x),(i \dels - m)  \psi(x)]_-   \nonumber          \\
& = & \sum_{\alpha=1}^p
( \bar{\psi}^{(\alpha)} (x)~ (i \dels - m) \psi^{(\alpha)}(x))
 + \mathrm{c-number}.                          \label{freepf}
\end{eqnarray}
The generating functional for a free charged parafermi field is
\begin{eqnarray}
& &\int \prod_{\alpha=1}^p {\cal D} \bar{\psi}^{(\alpha)\star} {\cal D}
\psi^{(\alpha)} ~\exp[-i \sum_1^p \int d^4x (
\bar{\psi}^{(\alpha)}(x) (i \dels - m)\psi^{(\alpha)}(x)  \nonumber   \\
&  & +\bar{\eta}^{(\alpha)}(x)\psi^{(\alpha)}(x)   +
 \bar{\psi}^{(\alpha)}(x)\eta^{(\alpha)}(x))]. \label{fgenfun}
\end{eqnarray}
Now we replace the path integral over commuting
fields by the Berezin path integral~\cite{ber} over Grassmann fields
and require
the Grassmann fields $\bar{\psi}^{(\alpha)}(x)$ and $\psi^{(\alpha)}(x)$ and
the
external sources $\bar{\eta}^{(\alpha)}(x)$ and $\eta^{(\alpha)}(x)$ to
anticommute for the same  and commute for
different values of the Green indices.  The generating functional
for  a free parafermi field is
\beee
W_{free}[\eta, \bar\eta]  = W_{free}[0,0] \exp \left[ - i \sum_{\alpha = 1}^p \int d^4x d^4y
~ \bar{\eta}^{(\alpha)} (x) S_F(x - y) \eta^{(\alpha)}(y)
\right].
~\\
~\\
~\\
~\\
\eeee           \label{genpf}
where $S_F(x)=(2\pi)^{-4}\int d^4p ~
\exp(-ip \cdot x) (\QQ + m) (p^2 - m^2 +i \epsilon)^{-1}$. In the presence
of an interaction $V(\bar{\psi}, \psi)$, we modify the generating functional
to
\beee
W[\eta, \bar\eta] =
\exp \left[i \int d^4x
V(-i \frac{\delta}{\delta \eta (x)}, -i \frac{\delta}{\delta \bar\eta(x)}) \right] W_{free}
[\eta, \bar\eta].
\eeee           \label{intpf}

These arguments can be extended further to systems in which
parafermi and parabose fields have mutual interaction.
As a specific example of such systems, we consider the
Yukawa coupling between a charged parafermi and a neutral parabose fields
of order 3, respectively \cite{owgmes}.  These fields obey relative para-Bose
commutation rules. The interaction term is
\begin{eqnarray}
{\cal L_I}  &=&
 g\frac{1}{4}[[\bar\psi(x),\psi(x)]_+
 - <[\bar\psi(x),\psi(x)]_+>_o,\phi(x)]_+ \nonumber    \\
 & =&   g\sum_{\stackrel{\alpha,\beta,\gamma}{\mathit{all ~different}}}
\bar\psi^{(\alpha)}(x)\psi^{(\beta)}(x)\phi^{(\gamma)}(x).
\end{eqnarray}
The generating functional for the noninteracting system is now
\begin{eqnarray}
\lefteqn{W_{free}[\eta, \bar\eta, J]   = }   \nonumber   \\
& & W_{free}[0,0,0] \exp \left[ - i \sum_{\alpha = 1}^3 \int d^4x d^4y
~ \bar{\eta}^{(\alpha)} (x) S_F(x - y) \eta^{(\alpha)}(y)
\right]                            \nonumber      \\
& \times &
\exp\left[-i \sum_{\gamma = 1}^3
\int d^4x d^4y J^{(\gamma)}(x) \Delta_F(x-y) J^{(\gamma)}(y)
\right]
\end{eqnarray}
which leads to the following expression for the complete
generating functional
\beee
W[\eta,\bar\eta, J]= \exp \left[- \int d^4x
\sum_{\stackrel{\alpha,\beta,\gamma}
{\mathit{all ~different}}}
\frac{\delta^3}{\eta^{(\alpha)}(x) \bar\eta^{(\beta)}(x)
J^{(\gamma)}(x)} \right]
W_{free}[\eta, \bar\eta, J].
\eeee

The propagators for the  parafermi system can be obtained by employing the
commutation relation Eq.(\ref{delta2}) with the lower sign. Note that
proceeding as in the parabose case, we get, for example, the following
expression for a four-point parafermi propagator
\begin{eqnarray}
\lefteqn{<0| T \psi (x_4) \psi (x_3) \bar{\psi}(x_2) \bar{\psi} (x_1) |0>  = }
\nonumber                            \\
& & \sum_{i_1,i_2,i_3,i_4} i^4
\frac {\displaystyle   \delta^4}{\displaystyle \delta \bar{\eta}^{(i_4)} (x_4) 
\delta \bar{\eta}^{(i_3)}(x_3)
 \delta \eta^{(i_2)} (x_2) \delta \eta^{(i_1)} (x_1)}
W|_{\eta = \bar{\eta}  = 0}     \nonumber    \\
&  =   & p^2 iS_F(x_3 - x_2) iS_F  (x_4 - x_1)
 -p(2-p)    iS_F  (x_3 - x_1)~  iS_F  (x_4 - x_2).      \label{4pointpf}
\end{eqnarray}
One can construct all other generating functionals for para fields, such
as the generating functionals for
connected and for one-particle-irreducible Greens functions,
in parallel with usual cases, just as we have done above for the time-ordered
products.

\section{Remark about the partition function}
The single-particle propagator is related to the partition
function~\cite{sch}. Our result Eq.(\ref{4point}) for 
the two-particle propagator gives a result for the 
partition function that is the square of the partition function in
the special case $p=2$ as we expect from the result of P. Suranyi~\cite{sur} that
a gas of $p=2$ parabosons (or parafermions) is equivalent to a mixture of
two independent gases of bosons (or fermions). We thank A. Polychronakos~\cite{pol1}
for pointing out that the different four-point matrix element that we gave in
an earlier version of this work
\begin{eqnarray}
\lefteqn{<0| T \phi^{\star} (x_4) \phi (x_3) \phi^{\star}(x_2) \phi (x_1) |0> =}
                                      \nonumber   \\
 &  & \sum_{\alpha,\beta,\gamma,\delta}   \left(
\frac {\displaystyle 1}{\displaystyle i  } \right)^4
\frac {\displaystyle   \delta^4}{\displaystyle \delta J^{\delta)} (x_4) \delta J^{(i_3)\star} (
x_3) \delta J^{(i_2)} (x_2) \delta J^{(i_1)\star} (x_1)} \times
W|_{J = 0}     \nonumber     \\
 &= &   p^2 [ i\Delta_F (x_1 - x_2) i\Delta_F (x_3 - x_4)
 +  i\Delta_F (x_1 - x_4) i\Delta_F (x_2 - x_3) ].
\end{eqnarray}          \label{4pointold}
does not have this property.
 
\section{Remark about normalization of para operators}
We have followed the normalization used by Green~\cite{gre} when he introduced
parastatistics in which the parameter $p$ that labels the order of the
parastatistics does not appear in the commutation relations~Eq.(\ref{paracr}), 
but $p$ does appear in the normalization
of the single-particle state~Eq.(\ref{norm}). Green's motivation was to keep the
usual commutation relations between the number operator and the annihilation
and creation operators~Eq.(\ref{number}). One can normalize the single-particle
state to one, but then the commutation relation becomes
\beee
[[b\dggg_k, b_l]_{\pm}, b\dggg_m]_-=2p\delta_{lm}b\dggg_k            \label{newparacr}
\eeee
and the number operator commutation relations also acquire the factor $p$.

\section{Comparison with earlier work}
We are aware of three earlier treatments of parastatistics using path
integrals. Y. Ohnuki and S.Kamefuchi~\cite{ohn} construct the path
integral for parafermi fields using paragrassmann variables
following the fundamental definition of
a path integral as the limit of a product of time evolution operators
for small time intervals. They do not evaluate this limit and thus do not
find an explicit formula for the path integral.
They also do not give the generating functions
that are most useful for calculations. Like our construction, their analysis
relies on the Green ansatz and does not give an intrinsic construction of
the path integral.
M. Chaichian and A.
Demichev~\cite{cha2} and A. Polychronakos~\cite{pol} both give an intrinsic
definition of the path integral that does not rely on the Green ansatz.
Their construction of the path integral is a first quantized analog of
using the Green
trilinear commutation relations without making use of the Green ansatz.
They define the propagator for paraparticles using the
propagator for distinguishable particles together with the matrix elements
of the permutation operator. They point out that these matrix elements are
``weights for each topological sector of the path integral'' that are
simply positive or negative integers or zero. The result of Polychronakos
that, in his notation, $S_p F(P) = (-1)^P S_pB(P)$, follows in our case from
the $\pm$ sign in Eq.(\ref{delta},\ref{delta2}).

\section{Summary and conclusions}
We have shown that parastatistics fields can be quantized using
path integrals. All the results are, as we expect, the same as found with 
the canonical formalism.
Up to now we have not found a computational advantage in using either of these
formalism for parastatistics. We explicitly calculated the two- and four-point
functions for free parabose and free parafermi theories and also
gave the
generating functionals for both free and interacting theories.
It would be interesting to formulate the path integral for parastatistics
fields using only a definition of the path integral that uses
the trilinear commutation relations for parafields without reference to
Green's ansatz.

\section{Acknowledgements}
We are happy to acknowledge the suggestion of John Tjon that stimulated us
to write this paper. We also greatly appreciate the suggestions of
John Tjon, Masud Chaichian and Alexios Polychronakos
that significantly improved our paper. We thank Cecile DeWitt for clarifying
the work of Laidlaw and DeWitt~\cite{lai}.
The work of OWG was supported in part by National
Science Foundation Grant No. PHY-0140301 and
in part by National Science Foundation Award No. INT-0223818.
The work of AKM was supported
in part by National Science Foundation Award No. INT-0223818 and by
the Department of Science and Technology of India.

\end{document}